\documentstyle[12pt]{article}
\newtheorem{definition}{Definition}
\newtheorem{theorem}{Theorem}
\newtheorem{proposition}{Proposition}
\newtheorem{example}{Example}
\newtheorem{lemdef}{Lemma-definition}

\begin{document}

\begin{center}
\parbox{15cm} {
\begin{center}
{\Large \bf On algebraic geometric and computer algebra aspects 
of mirror symmetry }
\end{center}
}
\end{center}

\begin{center}
{\large {\underline
             {N.M. Glazunov}
}
}
\end{center}

\begin{center}

\parbox{10cm} {
\begin{center}
 {\it
               Glushkov Institute of Cybernetics NAS,
\\
            03187 Glushkov prospect 40, Kiev-187 , Ukraine,
\\
               fax: (38 044) 266-7418,
\\
               e-mail:  glanm@yahoo.com}
\end{center} }
\end{center}

Abstract. 
{\footnotesize We survey some algebraic geometric aspects of mirror 
symmetry and duality in string theory. Some applications of computer 
algebra to algebraic geometry and string theory are shortly reviewed.\\}

\section{Introduction}
  This paper aims to be accessible to those with no previous experience 
in algebraic geometry; only some basic familiarity with linear and
polynomial algebra, group theory, topology and category theory will be
assumed.
  Symmetry principles always played an important role in mathematics
and physics. Development of theoretical physics in direction of string 
theory enlarged the context of symmetry considerations and included in 
it the notion of duality. String theory has following ingredients: 
(i) base space (open or closed string) $\Sigma$; (ii) target space $M$; 
(iii) fields: 
$X \rightarrow \Sigma \rightarrow M;$ (iv) action 
$S = \int {\cal L}(X,\varphi).$ where ${\cal L}$ is a 
Lagrangian~\cite{BG:STh}. Let $G$ be a group such that 
$G \supset SU(3) \times SU(2) \times U(1).$
Recall that if ${\cal L}(G\Phi) = {\cal L}(\Phi)$ then ${\cal L}$ is
$G-$invariant, or $G-$symmetry. 
In string theory~\cite{BG:STh} one of the beautiful symmetries is the 
radius symmetry $R \rightarrow  1/R$ of circle, known as 
$T-$duality~\cite{Di:L87,LVW:CR} and~\cite{GPR:TC} and references there in.
Authors of papers~\cite{CLS:SD,GP:NP} conjectured that a similar duality
might exist in the context of string propagation on Calabi-Yau (CY)
manifolds, where the role of the complex deformation on one
manifold get exchanged with the K\"{a}hler deformation on the dual
manifold. A pair of manifolds satisfying this symmetry is called
{\em mirror pair}, and this duality is called {\em mirror symmetry}.
From the point of view of physicists which did the remarkable
discovery  mirror symmetry is a type of duality that means that we 
may take two types of string theory and compactify them in two 
different ways and achieve "isomorphic" physics~\cite{Va:GPh}. Or in
the case of a pair of Calabi-Yau threefolds $ (X,Y)$ P. Aspinwall 
are said~\cite{As:ST} that $X$ and $Y$  
to be a {\em mirror pair} if and only if the type IIA 
string compactified on $X$ is "isomorphic to" the $E_{8} \times E_{8}$ 
heterotic string compactified on $Y.$ If in the case $X$ is 
Calabi-Yau threefold then $Y$ is the product of a $K3$ surface and 
elliptic curve and the following data specifies the theory~\cite{As:ST}. 
\\  
  1. A Ricci-flat metric on $Y.$ \\
  2. A $B-$field $\in H^{2}(K3 \times E,{\bf R}/{\bf Z}).$ \\
  3. A vector bundle $V \rightarrow (K3 \times E)$ with a connection
satisfying the Yang-Mills equations and whose structure group
$\subseteq E_{8} \times E_{8}.$ \\
  4. A dilation + axion, $\Phi \in {\bf C}.$ \\  
C. Vafa defines the notion of mirror of a Calabi-Yau 
manifold with a stable bundle. Lagrangian and special Lagrangian  
submanifolds appear in this situation. 
Mathematicians also work hard upon the problems of mirror symmetry,
although it is difficult in some cases to attribute to a researcher
the identifier "mathematician" or "physicist". 
V. Batyrev gives construction of mirror pairs using Gorenstein toric 
Fano varieties and Calabi-Yau hypersurfaces in these 
varieties~\cite{Ba:MS}.
M. Kontsevich in his talk at the ICM'94 gave a conjecturel 
interpratation of mirror symmetry as a "shadow" of an equivalence
between two triangulated categories associated with $A_{\infty}-$
categories~\cite{Ko:HA}. His conjecture was proved in the case of 
elliptic curves by A. Polishchuk and E. Zaslow~\cite{PZ:CM}. 
The aim of the paper is to provide a short and gentle survey of
some algebraic aspects of mirror symmetry and duality with examples - 
without proofs, but with (a very restricted) guides to the literature.

\section{Algebraic geometric preliminaries}
  This section gives a basic introduction to algebraic geometric aspects
of mirror symmetry beginning with a description of how Calabi-Yau
manifolds arise from ringed spaces. A more detailed treatment of 
algebraic geometric material of this section may be found 
in~\cite{Sh:FAG,GH:PAG}, the terminology and notation of which will be 
followed here.   \\
   Let $X$ be a topological space and $Cov(X)$ an open covering of $X.$
It is well known~\cite{Go:AT} that $Cov(X)$ forms the category. Let 
$Cat$ be a category (of sets, abelian groups, rings, modules). 
 The {\em presheaf} is a contravariant functor $\cal F$ from $Cov(X)$ to 
$Cat.$ If for example $Cat$ is the category $Ab$ of abelian groups then
${\cal F}: Cov(X) \rightarrow Ab$ is the presheaf of abelian groups. 
Elements $f \in \cal F(U)$ is called {\em sections} of the presheaf 
$\cal F.$    \\
If $i: U \subset V$ then we shall denote by $\rho_{U}^{V}$ the morphism
${\cal F}(i): {\cal F}(V) \rightarrow {\cal F}(U).$ Functor ${\cal F}$ 
is the contravariant and we apply morphisms from the left to the rigth.
Hence for any open sets  $U \subset V \subset W$  
$$ \rho_{U}^{W} = \rho_{U}^{V}\rho_{V}^{W}.$$  
Let $U \subset X$ be any open subset of $X$ and $\bigcup U_{\alpha} =
 U$ it's open covering. A presheaf ${\cal F}$ on a topological space
$X$ is called the $\em sheaf$ if the following conditions are 
satisfied: \\
1) if $ \rho_{U_{\alpha}}^{U} s_{1} = \rho_{U_{\alpha}}^{U} s_{2}$ for
$s_{1}, s_{2} \in {\cal F}(U)$ and for any $U_{\alpha},$ then 
$s_{1} = s_{2}.$ \\
2) if $s_{\alpha} \in {\cal F}(U_{\alpha})$ are such that 
$\rho_{U_{\alpha}\bigcap U_{\beta}}^{U_{\alpha}} s_{\alpha} = 
\rho_{U_{\alpha}\bigcap U_{\beta}}^{U_{\beta}} s_{\beta},$ then there
exists $s \in {\cal F}(U)$ such that $s_{\alpha} = 
\rho_{U_{\alpha}}^{U} s$ for all $U_{\alpha}.$

   The {\em ringed space} is the pair $(X,{\cal O}),$ where $X$ is a
topological space and $\cal O$ is a sheaf of rings on $X.$
A {\em morphism of ringed spaces} 
$$\varphi: \! (X,{\cal O}_{X}) \rightarrow (Y,{\cal O}_{Y})$$
is the class of maps $(\varphi,\psi_{U}),$ where $\varphi$ is the
continuous map $\varphi: \! X \rightarrow Y,$ and $\psi_{U}$ is a
homomorphism $\psi_{U}: \; {\cal O}_{Y}(U) \rightarrow 
{\cal O}_{X}(\varphi^{-1}(U))$ for any open set $U \subset Y$ such that
the diagram
$$
  \begin{array}{ccc}
 {\cal O}_{X}(\varphi^{-1}(V)) & \stackrel{\rho_{U}^{V}}{\longrightarrow}
 & {\cal O}_{X}(\varphi^{-1}(U))\\
 \downarrow\lefteqn{\psi_V}&&\downarrow\lefteqn{\psi_U}\\
  {\cal O}_{Y}(V)&\stackrel{\rho_{U}^{V}}{\longrightarrow} 
 & {\cal O}_{Y}(U)
  \end{array}
$$
 is commutative for any $U$ and $V, \; U \subset V \subset Y.$ 
For a ringed space $(X,{\cal O}_{X})$ and an open $U \subset X$ the
restriction of the sheaf ${\cal O}_{X}$ on $U$ defines the ringed space
$(U,{\cal O}_{X|U}).$ \\
Let ${\cal O}_{X}$ be the sheaf of smooth (differentiable) functions on 
a topological space $X.$ Then any smooth (differentiable) manifold $X$ 
with the sheaf ${\cal O}_{X}$ is a ringed space $(X,{\cal O}_{X}).$ 
   Respectively let $X$ be a Hausdorff topological space and 
${\cal O}_{X}$ a sheaf on $X.$ Let it satisfies conditions:
(i) ${\cal O}_{X}$ is the sheaf of algebras over $\bf C$~;
(ii) ${\cal O}_{X}$ is a subsheaf of the sheaf of continuous complex
valued functions. Let $W$ be a domain in ${\bf C}^n$ and 
${\cal O}_{an}$ the sheaf of analytical functions on $W.$ The
 ringed space $(X,{\cal O}_{X})$ is called the {\em complex analytical
manifold} if for any point $x \in X$ there exists a neighbourhood
$U \ni x$ such that $(U,{\cal O}_{X|U}) \simeq (W,{\cal O}_{an})$
(here $\simeq$ denotes the isomorphism of ringed 
spaces).        \\ 
   The {\it fibre space} is the object $(E,p,B)$, where $ p $
is the continuous surjective (= on) mapping of a topological
space $ E $ onto a space $ B $ (in our consideration $ B = M $
is a complex analytical manifold or algebraic variety), and 
$ p^{-1}(b) $ is called the
{\it fibre} above $ b \in B $. Both the notation
$ p:E \rightarrow B $
and $ (E,p,B) $ are used to denote a {\it fibration}, a
{\it fibre space}, a {\it fibre bundle} or a {\it bundle}. \\
{\it Vector bundle} is fibre space each fibre $ p^{-1}(b) $ of
which is endowed with the structure of a (finite dimension)
vector space $ V $ over skew-field $ K $ such that the following
local triviality condition is satisfied: each point $ b \in B $
has an open neighborhood $ U $ and a $ V $ -isomorphism of
fibre bundles $ \phi: \ p^{-1}(U) \rightarrow U \times V $ such
that $ \phi \mid_{p^{-1}(b)}: \ p^{-1}(b) \rightarrow b \times V $
is an isomorphism of vector spaces for each $ b \in B $. \\
 $ dim V $ is said to be the dimension of the vector bundle. \\
Let $X$ be a complex analytical manifold, $U$ an element of open
covering of $X, \; z_{1}, z_{2}, \cdots z_{n}$ coordinates on $U.$
Let $\varphi$ be the hermitian positively defined form on $X$,
$\varphi(z,\overline{z}) = 
\sum c_{\alpha,\beta}dz_{\alpha}d\overline{z}_{\beta}$ the hermitian
metric assosiated to $\varphi.$         
An {\em Hermitian bundle} over complex analytical manifold $X$
consists of a vector bundle over $X$ and a choice of $ C^\infty$
Hermitian metric on the vector bundle over complex manifold
$ X({\bf C}) $, which is invariant under antiholomorphic involution
of $ X({\bf C}) $.            \\
  Let $\varphi(z,\overline{z})$ be a hermitian metric on the tangent 
bundle on $X$ and $$ \omega = (i/2) \sum c_{\alpha,\beta}dz_{\alpha}
\wedge d\overline{z}_{\beta} $$
the two dimensional differential form assosiated to
$\varphi(z,\overline{z}).$ \\
The hermitian metric $\varphi$ on a complex analytical manifold $X$
is called {\em k\"{a}hlerian} if the differential form $ \omega$ 
assosiated to $\varphi$ is closed.  The {\em K\"{a}hler manifold}
is the complex analytical manifold with a k\"{a}hlerian metric.     
We shall use in contrast to~\cite{BG:STh} some another definition of 
Calabi-Yau (CY) manifold. The definition is based on the theorem of Yau 
who proved Calabi's conjecture that a complex K\"{a}hler manifold of 
vanishing first Chern class admits a Ricci-flat metric. 
\begin{definition}
 A complex K\"{a}hler manifold is called Calabi-Yau (CY) manifold if
it has vanishing first Chern class.
\end{definition}
  Examples of the CY-manifolds include, in particular, elliptic curves 
$E$, $K3-$surfaces and their products $E\times K3.$
 Let $(X,\omega,\Omega)$ be a complex manifold (real dimension =$2n$)
with 
$$\omega^{n}/n! = (-1)^{n(n-1)/2}(i/2)^{n} \cdot \Omega \wedge 
\overline{\Omega}.$$
It is said that a $n-$dimensional submanifold $L \subset X$ is
{\em special Lagrangian (s-lag)} $\Leftrightarrow$
$$ Re(\Omega|_{L}) = Vol|_{L} \; \Leftrightarrow
\omega|_{L} = 0, Im(\Omega|_{L}) = 0.$$
\begin{example}
Let $X$ be an elliptic curve $E.$ Then 
$ \omega = c(i/2)dz \wedge d\overline{z}, \; \Omega = cdz.$
S-lag $L \subset E$ are straight lines with slope determined by
$\arg c.$
\end{example}        
     Let $(U,{\cal O}_{X|U}) \simeq Spec A$ for a 
commutative ring $A$. In the case the neighbourhood $U \ni x$ is called 
the {\em affine neighbourhood} of the point $x.$ \\
   The {\em scheme}  $S$ is the ringed space $(X,{\cal O}_{X})$ with
the condition: for any point $x \in X$ there is an affine neighbourhood
$V \ni x$ such that $(V,{\cal O}_{X|V}) \simeq Spec A.$

\subsection{Blow-ups}

Blowing up is a well known method of constructing complex manifolds $M.$
There are points on the manifolds that are not divisors on $M.$ Blow up
is the construction that transforms points of complex manifolds to
divisors. For instance in the case of two dimensional complex manifolds
(complex surface) $N$ it consists of replacing a point $p \in N$ by a 
projective line ${\bf CP}(1)$ considered as the set of limit directions 
at $p.$
\begin{example}
 Let $\pi: M_{2} \rightarrow {\bf C}^2$ be the blow-up of ${\bf C}^2$
at the point $0 \in {\bf C}^2.$ Then $M_{2}$ is a two dimensional complex 
manifold that defined by two local charts. In coordinates
${\bf C}^2= (z_{1},z_{2}), {\bf CP}(1) = [l_{0},l_{1}]$ the manifold 
$M_{2}$ is defined in ${\bf CP}(1) \times {\bf C}^2$ by quadratic equations
$z_{i}l_{j} = z_{j}l_{i}.$ Thus $M_{2}$ is a line bundle over Riemann
sphere ${\bf CP}(1)$. $\pi^{-1}(0) = {\bf CP}(1)$ is called {\it the 
divisor of the blow up (the exceptional divisor)}.   
\end{example}

 Recently a large class of CY orbifolds in weighted projective spaces have
been proposed. C. Vafa have predicted and S. Roan~\cite{Ro:AG} have 
computed the Euler number of all the resolved CY hypersurfaces in
a weighted projective space ${\bf WCP}(4)$.

\subsection{Vector Bundles over Projective Algebraic Curves }

Let $X$ be a projective algebraic curve over algebraically
closed field $k$ and $g$ the genus of $X$. 
Let ${\cal VB}(X)$ be the category of vector bundles over 
$X$. Grothendieck have shown that for a rational curve every
vector bundle is a direct sum of line bundles. Atiyah have classified 
vector bundles over elliptic curves. The main result is~\cite{At:VB}: \\
\begin{theorem}
 Let $X$ be an elliptic curve, $A$ a
fixed base point on $X$. We may regard $X$ as an abelian
variety with $A$ as the zero element.  Let ${\cal E}(r,d)$ denote 
the the set of equivalence classes of indecomposable vector
bundles over $X$ of dimension $r$ and degree $d$. Then each
${\cal E}(r,d)$ may be identified with $X$ in such a way that \\
$ det: {\cal E}(r,d) \rightarrow {\cal E}(1,d) $ corresponds to 
$ H: X \rightarrow X, $ \\
where $ H(x) = hx = x + x + \cdots + x \; (h \:$ times$)$, and
$h = (r,d)$ is the highest common factor of $r$ and $d$.
\end{theorem}
Curve $X$ is called a {\it configuration} if its normalization
is a union of projective lines and all singular points of $X$
are simple nodes~\cite{DG:VB}. For each configuration $X$ can assign a
non-oriented graph $\Delta(X)$, whose vertices are irreducible
components of $X$, edges are its singular and an edge is
incident to a vertex if the corresponding component contains the
singular point. Drozd and Greuel have proved~\cite{DG:VB}: 
\begin{theorem}
 1. ${\cal VB}(X)$ contains finitely many
indecomposable objects up to shift and isomorphism if and only if
$X$ is a configuration and the graph $\Delta(X)$ is a simple chain
(possibly one point if $X = {\bf P}^1$). \\
2. ${\cal VB}(X)$ is tame, i.e. there exist at most one-parameter
families of indecomposable vector bundles over $X$, if and only if
either $X$ is a smooth elliptic curve or it is a configuration
and the graph $\Delta(X)$ is a simple cycle (possibly, one loop
if $X$ is a rational curve with only one simple node).   \\
3. Otherwise ${\cal VB}(X)$ is wild, i.e. for each finitely generated
$k-$algebra $\Lambda$ there exists a full embedding of the category
of finite dimensional $\Lambda-$modules into ${\cal VB}(X)$. 
\end{theorem} 
Let $X$ be an algebraic curve. How to normalize it? There are
several methods, algorithms and implementations for this 
purpose. A new algorithm and implementation is presented 
in~\cite{DJGP:NA}.

\section{Complexes, homotopy categories, cohomologies and
quasiisomorphisms}
  Here we recall the relevant properties of complexes, derived 
categories, cohomologies and quasimorphisms referring 
to~\cite{GH:PAG,GM:MH} for details and indication of proofs. \\
The {\em cochain complex} 
$$ (K^{\bullet},d) = \{K^{0} \stackrel{d}\rightarrow K^{1} 
\stackrel{d}\rightarrow K^{2} \stackrel{d}\rightarrow \cdots \}    
$$
is the sequence of abelian groups and differentials 
$d: K^{p} \rightarrow K^{p+1}$ with the condition $d \circ d = 0.$
For a category $Cat$ we denote by $Ob\, Cat$ it's objects and by
$Mor\, Cat$ it's morphisms.\\
Let $A$ be an abelian category, $Kom(A)$ the category of complexes over
$A$. Furthermore, there are various full subcategories of $Kom(A)$
whose respective objects are the complexes which are bounded below,
bounded above, bounded in both sides. Now recall (by~\cite{GM:MH}) the
notion of homotopy morphism.     

\begin{lemdef}
 (i) Let $K^{\bullet}, L^{\bullet}$ be two complexes over abelian category
$A$, $k = k^{i}, \; k^{i}: K^{i} \rightarrow L^{i-1}$ a sequence of
morphisms between elements of the complexes. \\
Then the maps
$$ h^{i} = k^{i+1}d_{K}^{i} + d_{L}^{i-1}k^{i}: K^{i} \rightarrow L^{i} $$
form the morphism of complexes
$$ h = kd + dk:  K^{\bullet} \rightarrow L^{\bullet}.$$
The morphism $h$ is called {\em homotopic} to zero ($h \sim 0$). \\
(ii) morphisms $f, g: K^{\bullet} \rightarrow L^{\bullet}$ is called 
homotopic, if $f - g = kd +dk \sim 0 \; (f \sim g)$, $k$ is called  
homotopy. \\
(iii) If $f \sim g,$ then $H^{\bullet}(f) = H^{\bullet}(g),$ where
the map $H^{\bullet}$ is induced on cohomologies of complexes.     
\end{lemdef}

 The {\em homotopy category} $K(A)$ is defined by the following
way: 
$$ Ob \; K(A) = Ob \; Kom(A), $$
$$ Mor \; K(A) = Mor\; Kom(A)\; \mbox{by the module of homotopy
equivalence}. $$ 
Let $X$ be a topological space and 
${\cal K}^{\bullet}, {\cal L}^{\bullet}$ be complexes of sheaves over
$X$. The {\em quasiisomorphism} is the map
$$ f: {\cal K}^{\bullet} \rightarrow {\cal L}^{\bullet}$$
which induces the isomorphism of cohomological sheaves
$$ f_{*}: {\cal H}^{q}({\cal K}^{\bullet}) \rightarrow
 {\cal H}^{q}({\cal L}^{\bullet}), \; q \geq 0.$$

\section {Connections}
  Consider the connection in the context of algebraic geometry.
Let $S/k$ be the smooth scheme over field $k$, $U$ an element of
open covering of $S$, $ {\cal O}_S $ the structure sheaf on $S$,
$\Gamma(U,{\cal O}_{S})$ the sections of ${\cal O}_{S}$ on $U$. 
Let $\Omega^{1}_{S/k}$ be the sheaf of germs of $1-$dimension 
differentials, $\cal F$ be a coherent sheaf. The {\em connection} on the 
sheaf $\cal F$ is the sheaf homomorphism 
$$ \nabla: {\cal F} \rightarrow \Omega^{1}_{S/k} \otimes {\cal F},$$
such that, if $ f \in \Gamma(U,{\cal O}_{S}), \;
g \in \Gamma(U,{\cal F})$ then
$$ \nabla(fg) = f\nabla(g) + df \otimes g.$$ 
  There is the dual definition. Let ${\cal F}$ be the locally
free sheaf, $\Theta^{1}_{S/k}$ the dual to sheaf $\Omega^{1}_{S/k}$, 
$\partial \in \Gamma(U,\Theta^{1}_{S/k})$. 
The {\em connection} is the homomorphism  
$$ \rho: \Theta^{1}_{S/k} \rightarrow 
End_{{\cal O}_S}({\cal F},{\cal F}).$$
$$\rho(\partial)(fg) = \partial(f)g + f\rho(\partial). $$ 

\subsection{Integration of connections}

Let $\Omega^{i}_{S/k}$ be the sheaf of germs of $i-$dimensional 
differential forms on $S.$ Particularly $\Omega^{1}_{S/k}$ is the
cotangent bundle over $S.$ Let $ \omega \in \Omega^{i}_{S/k}, \; f \in
\Gamma$ and
$$\nabla_{i}(\omega \otimes f) = d\omega \otimes f + (-1)^{i} \omega
\wedge \nabla (f).$$  
Hence, $\nabla_{i}$ define the sequence of homomorphisms of sheaves
$$ 
  \begin{array}{ccccccc}
{\cal F}& \stackrel{\nabla}{\longrightarrow} & \Omega^{1}_{S/k}\otimes
{\cal F} & \stackrel{\nabla_1}{\longrightarrow} & 
\Omega^{2}_{S/k}\otimes {\cal F} & \longrightarrow & \cdots
  \end{array}
$$
The sequence is the {\em complex} if $\nabla \circ \nabla_1 = 0.$ In 
this case the connection $\nabla$ is integrable. 
\begin{example}
  Let ${\cal F} = {\cal O}_S$ be the structural sheaf on $S.$ Then
$$\nabla: {\cal O}_{S} \rightarrow \Omega^{1}_{S/k}\otimes {\cal O}_{S}
\simeq \Omega^{1}_{S/k}.$$ 
Hence $\nabla(f) = df, \; \rho: \Theta_{S/k} \rightarrow {\cal O}_S.$ 
This connection $\nabla$ is integrable because it defines the de Rham 
complex 
$$ \Omega^{\bullet}_{S/k}: {\cal O}_{S} \rightarrow \Omega^{1}_{S/k}
\rightarrow  \Omega^{2}_{S/k} \rightarrow \cdots$$
\end{example}
\begin{example}
Let ${\cal LC}$ be a locally constant sheaf on $S/k$ such that ${\cal LC}
\simeq k^n$ (local coefficients) as sheaves. Let ${\cal F} = {\cal LC}
\otimes {\cal O}_{S}, \; v \in \Gamma(U,{\cal LC}), \; f \in 
\Gamma(U,{\cal F}).$ Then there is a {\em canonical connection} 
$\nabla(v\otimes f) = df \otimes v$:
$$\nabla: {\cal F} \rightarrow \Omega^{1}_{S} \otimes {\cal F}.$$   
\end{example}
For a connection $\nabla: {\cal F} \rightarrow \Omega^{1}_{S}\otimes 
{\cal F}$ a section $s \in \Gamma(U,{\cal F})$ is {\em horizontal} if
$\nabla(s) = 0.$ \\

Let now $S$ be a complex manifolds, $ShC(S/k)$ the category of 
sheaves with a connection and $LC(S)$ the category of local coefficients
over $S$. Let $({\cal F},\nabla) \in ShC(S/k)$ be a sheaf with
connection. We can define the functor 
$$Fn: ({\cal F},\nabla) \mapsto \{
\mbox{the sheaf of germs of horizontal sections of} \; {\cal F}\}.$$ 
and it's inverse $Fn^{-1}$. 
\begin{proposition}
The functors $Fn$ and $Fn^{-1}$ give the equivalence of category 
$ShC(S/k)$ and $LC(S).$  
\end{proposition}

\section{ Moduli spaces in string theory}

 Mirror symmetry connects with geometrical deformations of complex 
and K\"{a}hler structures on CY-manifolds. So we have to know moduli 
spaces of complex and K\"{a}hler structures on CY-manifolds.  
\subsection{Moduli spaces}
   The theory of moduli spaces~\cite{Ma:T,HM:MC} has, in recent
years, become the meeting ground of several different branches
of mathematics and physics - algebraic geometry, instantons, 
differential geometry, string theory and arithmetics. Here 
we recall some underlieing algebraic structures of the relation.
In previous section we have reminded the situation with vector
bundles on projective algebraic curves $X$. On $X$ any first Chern
class $c_{1} \in H^{2}(X,{\bf Z})$ can be realized as $c_1$ of 
vector bundle of prescribed rank (dimension) $r.$  How to classify
vector bundles over algebraic varieties of dimension more than $1?$
This is one of important problems of algebraic geometry and the
problem has closed connections with gauge theory in physics and 
differential geometry. Mamford~\cite{Ma:T} and others have 
formulated the problem about the determination of which cohomology
classes on a projective variety can be realized as Chern classes
of vector bundles?  Moduli spaces are appeared in the problem.
What is moduli? Classically Riemann claimed that $3g - 3$ (complex) 
parameters could be for Riemann surface of genus $g$ which would
determine its conformal structure (for elliptic curves, when
$g = 1,$ it is needs one parameter). From algebraic point of view 
we have the following problem: given some kind of variety,
classify the set of all varieties having something in common with
the given one (same numerical invariants of some kind, belonging
to a common algebraic family). For instance, for an elliptic
curve the invariant is the modular invariant of the elliptic
curve.  \\
Let {\bf B} be a class of objects. Let $ S $  be a scheme. 
A family of objects parametrized by
the $ S $ is the  set of objects

    $$ X_{s}: s \in S, X_{s} \in {\bf B} $$

 equipped with an additional structure compatible with the structure
  of the base $ S $.
Parameter varieties is a class of moduli spaces. These varieties
is very convenient tool for computer algebra investigation
of objects that parametrized by the parameter varieties. We
have used the approach for investigation of rational points
of hyperelliptic curves over prime finite fields~\cite{Gl:M01}.
\begin{example}
 Let $\omega_{1}, \: \omega_{2} \in {\bf C}, \; 
Im(\omega_{1}/\omega_{2}) >0,\; \Lambda = n\omega_{1} + m\omega_{2},\;
n,m \in {\bf Z} $ be a lattice. Let $H$ be the upper half plane.
Then $H/\Lambda = E$ be the elliptic curve. Let  
\[
 y^2 = x^3 + ax +b = (x-e_{1})(x-e_{2})(x-e_{3}) , \; 4a^3 + 27b^2 \neq 0,
\]
be the equation of $E.$ Then the differential of first kind on $E$
is defined by formula
\[
  \omega = dx/y = dx/(x^3 + ax +b)^{1/2}.
\]
Periods of $E:$
\[
 \pi_{1} = 2\int_{e_{1}}^{e_{2}} \omega,  \;       
 \pi_{2} = 2\int_{e_{2}}^{e_{3}} \omega.
\]
The space of moduli of elliptic curves over {\bf C} is 
${\bf A}^{1}({\bf C}).$ Its complation is ${\bf CP}(1).$  
\end{example}
 For $K3-$surfaces the situation is more complicated but in
some case is analogous~\cite{BPV:CC}:
\begin{theorem}\label{BPV:CC}  
 The moduli space of complex structure on market $K3-$surface
(including orbifold points) is given by the space of possible periods. 
\end{theorem}                
 Some computational aspects of periods and moduli spaces are
considered in author's note~\cite{Gl:C98}.   

\section{Some categorical constructions}

Every compact symplectic manifold $Y,\omega$ with vanishing first
Chern class, one can associate a $A_{\infty}-$category whose objects 
are essentially the Lagrangian submanifolds of $Y,$ and whose morphisms 
are determined by the intersections of pairs of submanifolds. This
category is called Fukaya's category and is denoted by 
${\cal F}(Y)$~\cite{Ko:HA}.  
Let $(X,Y)$ be a mirror pair. Let $M$ be any element of the mirror pair.
The bounded derived category $D^{b}(M)$ of coherent sheaves on $M$ is
obtained from the category of bounded complexes of coherent sheaves on 
$M$~\cite{GM:MH}. In the case of elliptic curves A. Poleshchuk and
E. Zaslov have proved~\cite{PZ:CM}:
\begin{theorem}\label{PZ:CM}  
 The categories $D^{b}(E_{q})$ and  ${\cal F}^{0}(\overline{E}^{q})$ 
are equivalent. 
\end{theorem}                
Recently A. Kapustin and D. Orlov have suggested that Kontsevich's
conjecture must be modified: coherent sheaves must be replaced with
modules over Azumaya algebras, and the Fukaya category must be 
"twisted" by closed 2-form~\cite{KO:RA}. 

\section{Computer algebra aspects}

Computer algebra applications to classical algebraic geometry are
well known. Most of them are based on the method of Gr{\"o}bner 
bases~\cite{Bu:AF,Bu:GB}. They include the decomposition of algebraic
varieties, rational parametrization of curves and 
surfaces~\cite{SW:SMC,Sc:RPS}, inversion of birational maps~\cite{Sc:IBM},
the normalization of affine rings~\cite{DJGP:NA}. Some computer algebra
results presented on CAAP-2001 can be used for computations in algebraic
geometry and string theory. These are results on computation of toric 
ideals~\cite{GB:JT} and on computation of 
cohomology~\cite{Ko:EM}. Talks of V. Gerdt include also result on 
computation in Yang-Mills mechanics~\cite{GKM:LC}.  
Some recent papers include description of efficient algorithms computing
the homology of commutative differential graded algebras~\cite{AARS:HP} 
and computing the complite Hopf algebra structure of the 1-homology of
purely quadratic algebra~\cite{JR:CS}.   \\
  For the future research let us mention that it might be as well to 
have a tool, namely computer algebra for computation with (i) various 
moduli spaces; (ii) deformations (deformation of complex structure and
deformation of K{\"a}hler structure); (iii) $A_{\infty}-$categories; 
(iv) geometric Fourier transform.

\subsection*{Conclusions}
 In the paper we tried to give an algebraic geometric framework for some 
aspects of mirror symmetry. This framework includes rather  restricted
context of mirror symmetry and string theory. But it is based on a
simple  and unified mathematical base. Some applications of computer 
algebra to algebraic geometry and string theory are shortly reviewed.

\subsection*{Acknowledgements}

  I would like to thanks the organizers of the CAAP'2001 (Dubna), 
SymmNMPh'2001 (Kiev) and SAGP'99 (Mirror Symmetry in String Theory, 
CIRM, Luminy)  for providing a very pleasent environment  during the 
conferences.

\pagebreak

\end{document}